# Comments on New Ontology of Quantum Mechanics called CSM

Marian Kupczynski

Département de l'Informatique, Université du Québec en Outaouais (UQO), Case postale 1250, succursale Hull, Gatineau. Quebec, J8X 3X 7, Canada

**Abstract** In this short article we show that a new quantum ontology of quantum mechanics proposed recently by Alexia Auffèves and Philippe Grangier is based on impossible to realize measurements which need to be performed repeatedly on the same single physical system or on the same pair of physical systems. We agree with Bohr and the authors that quantum mechanics is a contextual theory and that the experimental contexts have to be a part of any description of quantum phenomena but in our opinion their new ontology is neither convincing nor useful. In particular the authors claim that in spin polarization correlations experiments an outcome of Alice's experiment provides a distant context for the Bob's measurement to be performed and that this explains the peaceful coexistence between quantum mechanics and relativity. We show that, contrary to their claim, the authors are unable to explain why strong correlations, between the outcomes of distant local measurements, do exist and why they preserve a condition of parameter independence (non-signaling). Strangely enough the authors ignore that these strong but imperfect correlations can be explained in a local and causal way using statistical contextual interpretation of quantum mechanics what was demonstrated in several articles.

**Keywords** Bell Inequalities · Entanglement · Quantum Nonlocality · Quantum Measurement · Contextuality · Complementarity · Physical Reality · Statistical Interpretation

## 1 Introduction

The interpretation of quantum mechanics (QM) is still a subject of animated discussions. Authors are trying to explain in intuitive way: wave packet reduction or its absence, wave-particle duality, quantum nonlocality and what is going behind the scenes. The believers in a particular interpretation dismiss all other interpretations with nearly religious fervor and refuse to read and understand what does not comply with their beliefs

In older literature on QM and in some recent articles on quantum information one may find a statement that a quantum pure state ψ provides a complete and exhaustive description of a single individual physical system. Several arguments why this statement is not only unproven but incorrect may be found in [1].

According to modern statistical and contextual interpretation (SCI), free of paradoxes, a pure state describes the statistical properties of an ensemble of similarly prepared physical systems [1-8]. In conformity with Bohr's conviction that there is no Quantum World but only an abstract quantum mechanical description of quantum phenomena SCI does not assume any quantum ontology. The question whether quantum mechanics may emerge from a more detailed theory, to be discovered, describing what is going behind the scenes is left open.

In SCI quantum states pure and mixed are only mathematical entities which together with Hermitian operators representing different physical observables and positive-operator valued measures (POVM's ) are used to explain in quantitative way various phenomena and also to find probabilistic predictions for the statistical spread of outcomes from repeated measurements performed on the ensemble of identically prepared physical systems in well-defined experimental contexts.

The contexts are defined by macroscopic instruments which produce outcomes to be analyzed by on-line computers and researchers. In SCI different experimental contexts provide different contextual information what is consistent with Bohr's ides of complementarity according to which mutually exclusive though complementary experiments embrace as a whole, everything, which can be <u>experienced</u> with regard to studied physical systems.

Let us underline that according to SCI the wave function reduction is not instantaneous and a reduced quantum state represents only a sub-ensemble of the initial ensemble of physical systems prepared or selected in a particular way. In SCI quantum probabilities are objective characteristics of quantum phenomena [8] and not as in Qbism subjective beliefs of some human agents.

After this introduction let us now analyze the article in which Alexia Auffèves and Philippe Grangier [9] try to introduce a new quantum ontology and claim that it explains various quantum paradoxes.

## 2  New  Quantum Ontology: Contexts, Systems and Modalities

In 2001 Philippe Grangier [10] postulated that " contextual objective reality" should be given to the quantum state of the physical system.  He defines the quantum state as follows: *The quantum state of a physical system is defined by the values of a set of physical quantities, which can be predicted with certainty and measured repeatedly without perturbing in any way the system. This set of quantities must be complete in the sense that the value of any other quantity which satisfies the same criteria is a function of these values*. He gives also an example of a physical system behaving according to his definition: *What actually gives a certain and reproducible result is a joint  Bell measurement on the pair of particles*.

As we see the author speaks about a single physical system or about a single pair of physical systems.  It is simply impossible to make measurements repeatedly, on the <u>same</u> single quantum system , without perturbing it in any way, Therefore the definition given above is meaningless. QM gives no prediction for a single outcome of a measurement before it is done. The main discovery of QM was that the outcomes of measurements of physical observables are not predetermined but they are obtained   in the interaction of  physical systems with  measuring instruments. Only macroscopic single physical system can observed repeatedly and only for them measurement outcomes can be predicted with quasi certainty.

In their recent paper Alexia Affèves and Philippe Grangier [9] define their new ontology (CSM) based on three notions "Context, System and Modality".

- A <u>single physical system</u> is defined as a subpart of the world that is isolated well enough to be studied.
- An <u>environment</u> in which a given set of questions can be asked to the system is called a <u>context</u>.
- A set <u>of answers which can be predicted with certainty and obtained repeatedly</u> within a particular context is called modality.

The contexts are classical. Modalities are attributed jointly to the system and the context and because they can be predicted with certainty and obtained repeatedly they fulfill the condition for the objective definition of a quantum state given in [10] and cited above.

As the first example the authors give a photon passing by different $\theta$-oriented polarizers. Unfortunately the experiments with a single photon they discuss cannot be done. We describe below the experiments which can be done and the information which can be obtained.

Let us consider a "stable" laser producing a beam $b_0$ of "photons". We can measure its intensity $I_0$ in counting *n* clicks registered by a photon detector during some fixed time window t: $I_0 = n/t$. Of course if we repeat our experiment we will get in general different results $I_0(i)$ scattered around some average intensity $<I_0>$ characterizing a source. Let us study properties of our beam using three different experiments (contexts).

In the first context $C_1$ we project our beam on $\theta$-oriented polarizer $P_1$ and we measure the intensity $I_1$ of the transmitted beam $b_1$ using the same time window as before. By repeating our experiment many times we obtain several results $I_1(j)$ scattered around some average intensity $<I_1>$. The ratio $R_{10}=<I_1>/<I_0>$ and its statistical error is a characteristic of "the beam $b_0$ and the context $C_1$".

In the second context $C_2$ we project the beam $b_1$ on a second $\theta$-oriented polarizer $P_2$, obtaining a beam $b_2$, and we measure its intensity $I_2$. By repeating our experiment many times we obtain several results $I_2(k)$ scattered around some average intensity $<I_2>$. The ratio $R_{21}=<I_2>/<I_1>$ and its statistical error is a characteristic of " the beam $b_1$ and the context $C_2$" or in other words is a characteristic of a whole experiment: a laser beam which passed by a $\theta$-oriented polarizer is analyzed by a second $\theta$-oriented polarizer.

In CSI we say that the beam $b_1$ is prepared in a pure quantum state. In spite of the fact that $R_{21} \approx 1$ ($R_{21} \neq 1$) we cannot say that for <u>any given photon</u> a passage by the polarizer $P_2$ <u>can be predicted with certainty and obtained repeatedly.</u> As Bohr underlined several times QM describes a quantum phenomenon as a whole and does not provide a model how the successive experimental outcomes are produced.

In the third context $C_3$ we project the beam $b_1$ on a $\theta$'-oriented polarizer $P_3$ obtaining a beam $b_3$ and we measure its intensity $I_3$. By repeating our experiment many times we obtain several results $I_3(m)$ scattered around some average intensity $<I_3>$. The ratio $R_{31}=<I_3>/<I_1>$ and its statistical error is a characteristic of " the beam $b_1$ and the context $C_3$" or in other words is a characteristic of whole experiment: a laser beam which passed by a $\theta$-oriented polarizer $P_1$ is analyzed by a $\theta$'-oriented polarizer $P_3$. According to Malus law and QM the ratio $R_{31} \approx \cos^2(\theta - \theta')$.

This ratio characterizes transmission properties of two macroscopic polarizers when the electromagnetic wave or a laser beam is projected on them and the signal intensity is evaluated classically or by counting the "photons" (clicks) registered by detectors. A more detailed discussion of similar experiments may be found in [7, 12].

## 3 Strong Correlations and Bell Inequalities

The authors claim that CSM allows better understanding of spin polarization correlation experiments (SPCE). They consider *two spin ½ particles in the singlet state shared between Alice and Bob. The singlet state is a modality among four mutually exclusive modalities defined in the context for two spins, where measurements of the total spin (and any components of this spin) will certainly and repeatedly give a zero value.* Apparently we should not be surprised by this statement because similar phrasing is used in many articles on quantum information. However this phrasing is incorrect.

There is no experimental capability allowing, in a <u>single measurement</u> performed on two *spin ½ particles* or two photons, to decide that this particular pair was indeed prepared in a spin singlet state. The reconstruction of a quantum state is a difficult and often not unambiguous task. In order to get information in which quantum state these physical systems were prepared we have to make the quantum state tomography based on a large number of outcomes from various experiments performed on the ensemble of identically prepared physical systems. Therefore a singlet state is not a joint modality of a <u>single</u> pair of EPR particles and of a corresponding experimental context.

We agree with the authors of [9] that the violation of Bell inequalities in several experiments proved that it is impossible to attribute predetermined properties to Bob's and Alice's particle when the pairs are prepared in a spin singlet state. However we disagree with the statement that: *"Alice can predict with certainty the state of Bob's particle; however, certainty applies jointly to the new context (owned by Alice) and to the new system (owned by Bob)."*

As we pointed several times [5,6] the orientation of a polarizer is not defined by a real number $\theta$ but only by a small interval $[\theta-\Delta\theta, \theta+\Delta\theta]$ thus QM does not predict strict anti-correlations in SPCE. Besides in order to reproduce experimental results of SPCE instead of a spin singlet state one has to use much more complicated quantum states [13]. Therefore statements that Alice's outcome $a_i$ is a new context for a distant Bob's particle and that: *if Bob does a measurement in the same context as Alice, he will find will certainty a result opposite to Alice's one* are incorrect.

The explanation of strong correlations in SPCE, given by CSM, is not only artificial but insufficient. The authors write an equation

$$p(a_i, b_j | \mu) = p(a_i | \mu) p(b_j | \mu, a_i) = p(b_j | \mu) p(a_i | \mu, b_j). \tag{1}$$

where $\mu$ denotes a singlet state and ($a_i$, $b_j$) Alice's and Bob's outcomes. They say that ($\mu$, $a_i$) defines a modality for Bob's particle by using only Alice's data.

Please note that in SPCE we may estimate only $p(a_i, b_j | \mu)$, $p(a_i | \mu)$ and $p(b_j | \mu)$. Therefore (1) is simply a standard definition of conditional probabilities $p(b_j | \mu, a_i)$ and $p(a_i | \mu, b_j)$. This definition cannot explain why the outcomes, obtained locally in distant laboratories, respect *the parameter independence* (non-signalling) and why they are more strongly correlated that it is permitted by local realistic hidden variable models (LRHV) or stochastic hidden variable models (SHV).

Much simpler explanation of long range correlations is given by SCI according to which $p(b_j | \mu, a_i)$ is only a probabilistic prediction for the statistical spread of outcomes obtained in measurements performed on a sub-ensemble of Bob's particles which are partners of Alice's particles for which measurements performed by Alice gave the outcome $a_i$.

It has been shown by several authors that LRHV and SHV neglect contextual character of quantum observables. All the proofs of Bell–type inequalities are based on counterfactual reasoning or/and on incorrect probabilistic models [3-7, 14-29]. It is surprising that the authors of [9] who underline the importance of contexts in QM ignore the existence of these articles.

In order to obtain a contextual and causally local description of SPCE one may consider a following probabilistic *contextual hidden variable* model. We have a source producing two correlated signals $S_1$ and $S_2$ sent to distant laboratories x and y where the outcomes $a_i = \pm 1$ or 0 and $b_j = \pm 1$ or 0 are produced in synchronized time-windows. If we skip 0 outcomes corresponding to the lack of a click there are 4 possible outcomes for ($a_i$, $b_j$).

We introduce variables $\lambda_1 \in \Lambda_1$, $\lambda_2 \in \Lambda_2$ and $P(\lambda_1, \lambda_2)$ describing the signals in the moment of the measurement and variables $\lambda_x \in \Lambda_x$, $\lambda_y \in \Lambda_y$, $P_x(\lambda_x)$ and $P_y(\lambda_y)$ describing measuring devices ( <u>as they are perceived</u> by incoming signals).

To preserve a partial memory of correlations created by a source outcomes $a_i$ and $b_j$ are produced in a local and deterministic way $a_i = A_x(\lambda_1, \lambda_x)$ and $b_j = B_y(\lambda_2, \lambda_y)$ where $A_x$ and $B_y$ are functions equal $\pm 1$. Thus the conditional probabilities:

$$p(a_i, b_j \mid \mu, x, y) = \int_{\Lambda_{xy}} P(\lambda) \delta(A_x(\lambda_1, \lambda_x) - a_i) \delta(B_y(\lambda_2, \lambda_y) - b_j) d\lambda_1 d\lambda_2 d\lambda_x d\lambda_y \qquad (2)$$

where $P(\lambda) = P(\lambda_1, \lambda_2) P_x(\lambda_x) P_y(\lambda_y)$ and $\Lambda_{xy} = \Lambda_1 \times \Lambda_2 \times \Lambda_x \times \Lambda_y$ depend explicitly on the experimental context $(x, y)$. A simpler formula for expectation values $E(A,B|x,y)$, when discrete supplementary parameters are used, may be found in [7].

It is obvious that Bell type inequalities cannot be proven using (2) and one has enough flexibility to fit any post-selected non zero outcomes from SPCE obtained in a given randomly chosen experimental setting $(x, y)$. We see that probabilistic model (2) satisfies *parameter independence* since a summation over all possible values of bj leads to marginal probability distributions: $p(a_i \mid \mu, x, y) = p(a_i \mid \mu, x)$ and $p(b_j \mid \mu, x, y) = p(b_j \mid \mu, y)$.

Therefore we see that SCI supplemented by a contextual causal local probabilistic model (2) is able to reproduce the predictions of QM and explain the violation of Bell-type inequalities in a much more convincing way than CSM.

## 4 Conclusions

In spite of the fact that we find new quantum ontology introduced by CSM neither convincing nor necessary we agree [8] with several statements found in [9].

Namely we agree that:

- *The goal of physics is to study entities of the natural world, existing independently from any particular observer's perception and obeying universal and intelligible rules.*
- *There exist an "ultimate reality", constituted by all the objects in nature which are, from a scientific point of view, made of particles, waves and all their combinations, giving rise to macroscopic bodies.*
- *Physics always deals with "empirical reality". Its duty is described phenomena with mathematical tools, which will allow one to predict the values of measurable physical quantities.*
- *It is very difficult to speak about anything like a "quantum state of universe"* [10].

Sets of axioms enumerated in older textbooks on QM are rooted in the experiments which can be done in quantum optics. Experimental capabilities are in general much more limited. For example in nuclear and in elementary particle physics the information obtained is based on collisions of carefully prepared beams with various targets and on a study of the effects caused by these collisions.

Strict localisation of an elementary particle is impossible. An attempt to localize an elementary particle would, in general, destroy this particle and would produce several different particles. Therefore particle positions are not used in high energy physics.

Instead momenta of charged particles emerging from collisions can be measured for example in drift chambers. The energy, also of neutral particles, may be measured in various calorimeters. The total linear momentum and energy conservation laws are used extensively. Various new conservation or partial conservation laws characterising different type of interactions are discovered and applied. The relation $\Delta E \Delta t \approx \Delta E \Delta t \approx \hbar/2$ is extensively used to estimate the life time of resonance particles.

Theoreticians predict existence of new particles and resonances which are discovered in the experiments. Therefore we have an experimental support to believe in the "ontological existence" of these physical systems.

Quantum mechanics and quantum field theory provide only abstract mathematical tools allowing obtaining quantitative predictions without giving an intuitive and detailed description of quantum phenomena. Nevertheless particle physicists are talking about *coloured quarks exchanging gluons* which are bounded inside hadrons due to the *infrared slavery*. In high energy collisions hadrons are described by *generalized parton distribution functions*. *Parton-parton* interactions are described by quantum chromodynamics and several produced quark –antiquark pairs and gluons *recombine* in the process of *hadronization* to form final particles.

Strangely enough these *partly intuitive inexact images* help to create new better mathematical models and to make new experimental discoveries. This impressive progress in the domain of particle physics gives no indication that the Nature is nonlocal thus, as we have shown also above, the claim that the violation of Bell–type inequalities proves that there is a new law of Nature called a *nonlocal randomness* is completely unjustified [29].

Let us finish our article by citing two passages from Max Born's Noble Lecture [30]:

- *Can we call something with which the concepts of position and motion cannot be associated in the usual way, a thing, or a particle?... I am emphatically in favor of the retention of the particle idea… Every object that we perceive appears in innumerable aspects. The concept of the object is the invariant of all these aspects. From this point of view, the present universally used system of concepts in which particles and waves appear simultaneously, can be completely justified."*
- *The latest research on nuclei and elementary particles has led us, however, to limits beyond which this system of concepts itself does not appear to suffice. The lesson to be learned from what I have told of the origin of quantum mechanics is that probable refinements of mathematical methods will not suffice to produce a satisfactory theory, but that somewhere in our doctrine is hidden a concept, unjustified by experience, which we must eliminate to open up the road.*

# References

1. Ballentine, L. E.: Quantum Mechanics: A Modern Development . World Scientific ,Singapore(1998)
2. Allahverdyan, A. E., Balian, R. and Nieuwenhuizen, T. M.: Understanding quantum measurement from the solution of dynamical models. Physics Reports 525, 1-166 (2013)
3. Khrennikov, A. Yu.: Interpretation of Probability. VSP,Utrecht (1999)
4. Khrennikov, A. Yu.: Contextual Approach to Quantum Formalism . Springer, Dortrecht (2009)
5. Kupczynski, M.: Seventy years of the EPR paradox. AIP Conf. Proc. 861, 516-523 (2006)
6. Kupczynski M.: EPR paradox ,locality and completeness of quantum theory. AIP Conf. Proc. 962, 274-85 (2007)
7. Kupczynski M: Bell inequalities, experimental protocols and contextuality. Found. of Phys. 45, 735 (2015) doi:10.1007/s10701-014-9863-4FOP
8. Kupczynski M.,: Greeks were right : critical comments on Qbism. arXiv:1505.06348v2 [quant-ph](2015)
9. Auffèves A. and Grangier P.: Contexts, systems and modalities: a new ontology for quantum mechanics. Found. of Phys. http://dx.doi.org/10.1007/s10701-015-9952-z (2015)
10. Grangier.P.: Contextual objectivity : a realistic interpretation of quantum mechanics. European Journal of P 23, 331 (2002)
11. Allahverdyan, A. E., Balian, R. and Nieuwenhuizen, T. M.: Understanding quantum measurement from the solution of dynamical models. Physics Reports 525, 1-166 (2013)
12. Kupczynski, M.: Is Hilbert space language too rich. Int.J.Theor.Phys.79, 319-43(1973), reprinted in: Hooker, C.A (ed).Physical Theory as Logico-Operational Structure, 89-113. Reidel,Dordrecht (1978)
13. Kofler J., Ramelow S., Giustina M. and Zeilinger A.: On Bell violation using entangled photons without the fair-sampling assumption. arXiv:1307.6475 [quant-ph] (2013)
14. Fine A.: Joint distributions, quantum correlations and commuting observables. JMP 23, 1306 (1982)
15. Kupczynski. M.: Bertrand's paradox and Bell's inequalities. Phys.Lett. A 121,205-07( 1987)
16. De Muynck, V. M. , De Baere, W. and Martens, H. : Interpretations of quantum mechanics, joint measurement of incompatible observables and counterfactual definiteness. Found. Phys. 24 1589-664 (1994)
17. De Muynck,W.M.:Foundations of Quantum Mechanics .Kluver Academic, Dordrecht (2002)
18. Nieuwenhuizen, T. M.: Where Bell went wrong. AIP Conf. Proc. 1101, 127-33 (2009)
19. Nieuwenhuizen, T. M.: Is the contextuality loophole fatal for the derivation of Bell inequalities. Found. Phys. 41, 580 (2011)
20. Hess K. and Philipp W.: Bell's theorem:critique of proofs with and without inequalities. AIP Conf. Proc. 750, 150 (2005)
21. Hess K., Michielsen K. and De Raedt H.: Possible Experience: from Boole to Bell. Europhys. Lett. 87, 60007 (2009)
22. Hess K., De Raedt H. and Michielsen K.: Hidden assumptions in the derivation of the theorem of Bell. Phys. Scr. T151, 014002 (2012)
23. Khrennikov A. Yu.: Bell's inequality: nonlocality, "death of reality", or incompatibility of random variables. AIP 962, 121(2007)
24. Khrennikov, A. Yu. Violation of Bell's inequality and nonKolmogorovness. AIP Conf. Proc. 1101, 86 (2009)
25. Khrennikov, A. Yu.: Bell's inequality: physics meets probability. Information Science 179, 492-504 (2009)
26. Khrennikov, A. Yu.: CHSH inequality: Quantum probabilities as classical conditional probabilities. Found. of Phys. 45, 711(2015)
27. Żukowski,M and Brukner, Č .: Quantum non-locality—it ain't necessarily so. J. Phys. A: Math. Theor. 47 (2014) 424009 (10pp). doi:10.1088/1751-8113/47/42/424009
28. Kupczynski, M.: Causality and local determinism versus quantum nonlocality. J. Phys.: Conf. Ser. 504 012015(2014). doi:10.1088/1742-6596/504/1/012015
29. Kupczynski, M.: EPR Paradox, Quantum Nonlocality and Physical Reality. arXiv:1602.02959 [quant-ph] (2016)
30. Born M.: The statistical interpretation of quantum mechanics www.nobelprize.org/nobel_prizes/physics/laureates/.../born-lecture.pdf (1954)